\documentclass[conference]{IEEEtran}
\IEEEoverridecommandlockouts
\usepackage{cite}
\usepackage{amsmath,amssymb,amsfonts}
\usepackage{amsmath}
\usepackage{cuted}
\usepackage{breqn}
\usepackage{graphicx}
\usepackage{textcomp}
\usepackage{xcolor}
\usepackage{mathtools}
\usepackage[linesnumbered,algoruled]{algorithm2e}
\usepackage{algpseudocode}
\usepackage{amsthm}
\usepackage{comment}
\usepackage{bm}
\usepackage{wrapfig}
\usepackage{chemformula}
\usepackage{balance}
\usepackage[free-standing-units=true]{siunitx}

\usepackage[all]{background}

\usepackage{stackengine}
\setstackEOL{\\}
\setstackgap{L}{\normalbaselineskip}
\SetBgContents{\color{gray}{\tiny \Longstack{PREPRINT - accepted by 21st IEEE Interregional NEWCAS Conference - An IEEE CAS Society Interregional Flagship Conference}}}% Set contents
\SetBgPosition{4.5cm,1cm}% Select location
\SetBgOpacity{1.0}% Select opacity
\SetBgAngle{0}% Select rotation of logo
\SetBgScale{1.8}% Select scale factor of logo

\begin{document}
\bstctlcite{IEEEexample:BSTcontrol}
\title{Gate Camouflaging Using Reconfigurable ISFET-Based Threshold Voltage Defined Logic \vspace{-5mm}}
% Animal: Anti-ML Locking for RTL
\author{\IEEEauthorblockN{Elmira Moussavi\IEEEauthorrefmark{1},
Animesh Singh\IEEEauthorrefmark{1},
Dominik Sisejkovic\IEEEauthorrefmark{2},
Aravind Padma Kumar\IEEEauthorrefmark{1},\\
Daniyar Kizatov\IEEEauthorrefmark{1},
Sven Ingebrandt\IEEEauthorrefmark{1},
Rainer Leupers\IEEEauthorrefmark{1}, 
Vivek Pachauri\IEEEauthorrefmark{1}, and
Farhad Merchant\IEEEauthorrefmark{3}
}
		\IEEEauthorblockA{\IEEEauthorrefmark{1} RWTH Aachen University, Germany \\ \IEEEauthorrefmark{2} Security, Privacy, and Safety Research Group, Corporate Research Robert Bosch GmbH, Germany\\ \IEEEauthorrefmark{3} Newcastle University, UK}
        %\IEEEauthorblockA{RWTH Aachen University, Germany, Newcastle University, UK}
        \{moussavi, padma.kumar, kizatov, leupers\}@ice.rwth-aachen.de, dominik.sisejkovic@de.bosch.com \\ \{singh, ingebrandt, pachauri\}@iwe1.rwth-aachen.de,  farhad.merchant@newcastle.ac.uk \vspace{-4mm}
        }

%\IEEEoverridecommandlockouts \IEEEpubid{\makebox[\columnwidth]{978-1-6654-5707-1/22/\/\$31.00 ~\copyright2022 IEEE \hfill} \hspace{\columnsep}\makebox[\columnwidth]{ }}
\maketitle
\IEEEpubidadjcol

\begin{abstract}
Most chip designers outsource the manufacturing of their integrated circuits (ICs) to external foundries due to the exorbitant cost and complexity of the process. This involvement of untrustworthy, external entities opens the door to major security threats, such as reverse engineering (RE). RE can reveal the physical structure and functionality of intellectual property (IP) and ICs, leading to IP theft, counterfeiting, and other misuses. The concept of the threshold voltage-defined (TVD) logic family is a potential mechanism to obfuscate and protect the design and prevent RE. However, it addresses post-fabrication RE issues, and it has been shown that dopant profiling techniques can be used to determine the threshold voltage of the transistor and break the obfuscation. In this work, we propose a novel TVD modulation with ion-sensitive field-effect transistors (ISFETs) to protect the IC from RE and IP piracy.  Compared to the conventional TVD logic family, ISFET-TVD allows post-manufacture programming. The ISFET-TVD logic gate can be reconfigured after fabrication, maintaining an exact schematic architecture with an identical layout for all types of logic gates, and thus overcoming the shortcomings of the classic TVD. The threshold voltage of the ISFETs can be adjusted after fabrication by changing the ion concentration of the material in contact with the ion-sensitive gate of the transistor, depending on the Boolean functionality. The ISFET is CMOS compatible, and therefore implemented on 45 nm CMOS technology for demonstration.
\end{abstract}

\section{Introduction}
Building and maintaining a semiconductor foundry is a challenging and costly process. Increased demand for chips and the competition for time to market has led many chip designers to outsource the fabrication to third-party foundries~\cite{vsivsejkovic2020scaling}. This acts as an enabler for security threats such as reverse engineering~\cite{fyrbiak2017hardware}, leading to the theft of the design's valuable intellectual property (IP piracy)~\cite{rostami2014primer}, and counterfeiting~\cite{guin2014counterfeit}. In RE-based techniques, the adversary de-layers the IC to obtain information about the functionality of the gates and their wire connectivity to reconstruct the netlist~\cite{de2017preventing}. 
To identify the gate's functionality and internal structure for post-manufacturing RE, techniques such as probing and high-resolution imaging are used to exploit critical information from the chip to perform various attacks efficiently~\cite{7495587}\cite{rai2021vertical}. 
 
\begin{figure}[!t]
   \centering
   \includegraphics*[width=0.8\columnwidth]{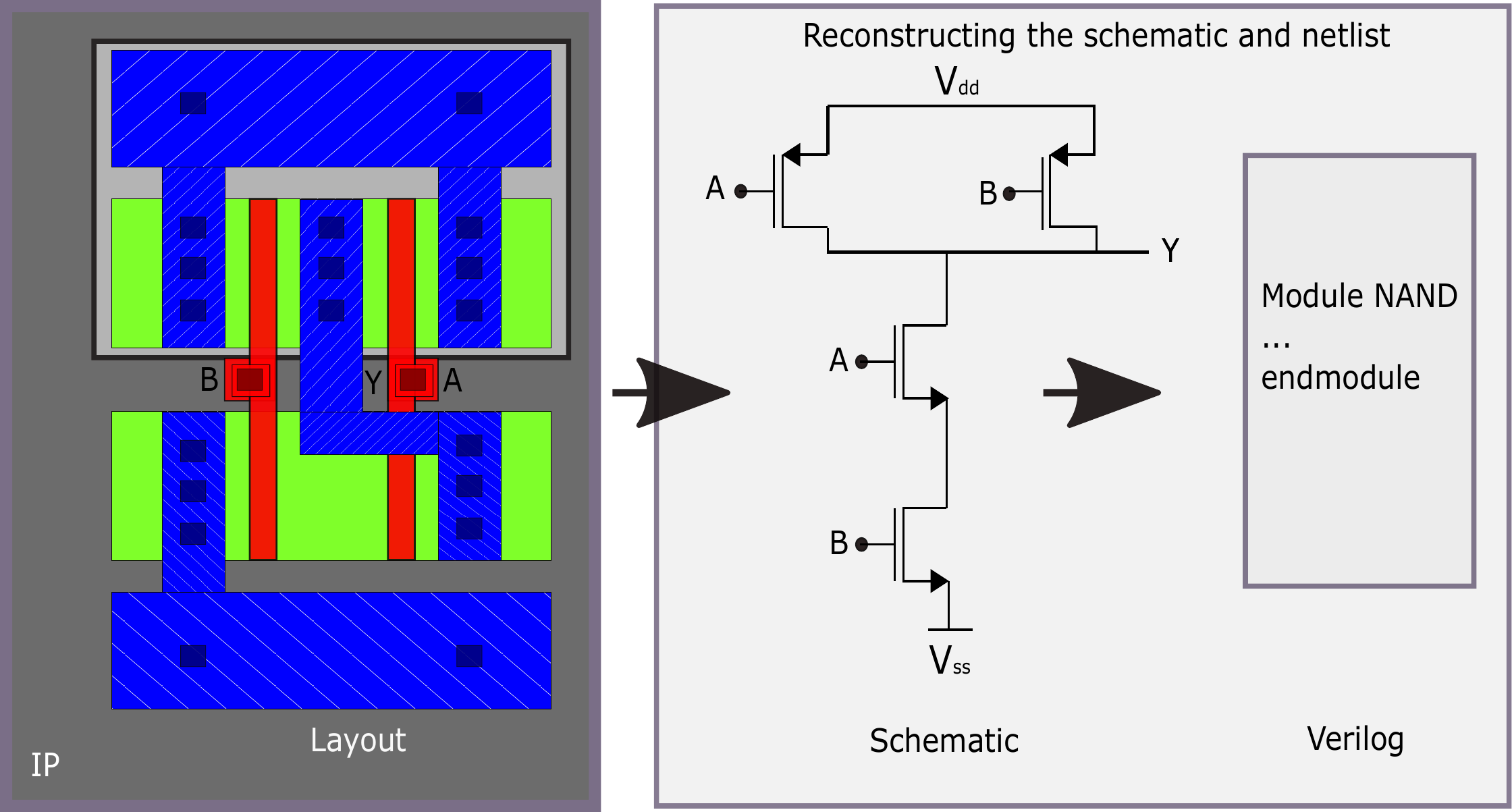}
   \caption{Example of a standard NAND gate that can be easily identified by looking at the top metal layers, reverse engineering, and reconstructing the design to clone the IP.}
   \label{fig:NAND}
   \vspace{-5mm}
\end{figure}

After de-packaging the IC, the attacker captures images of the metal and base layers, which contain information about the metal connections used for device interconnection and gate identification, respectively. By compiling the information obtained from the images, the attacker can eventually reconstruct the netlist for IP overproduction and illegally sell the cloned design on the black market (Fig.~\ref{fig:NAND}). 
Researchers have proposed various camouflaging techniques to combat RE. One such countermeasure is gate camouflaging. The primary purpose of gate camouflaging is to hide the functionality of particular logic gates and make RE impossible or very difficult by solely observing physical characteristics.

The threshold voltage-defined (TVD) logic family is a gate camouflaging technique to protect the design against post-manufacturing RE. TVD uses a combination of transistors with different threshold voltages ($V_{th}$) to define a gate functionality within \textit{an identical layout}. The procedure requires a one-time mask that is programmed with different threshold implants depending on the intended Boolean functionality. Consequently, to understand the gates' functionality, the $V_{th}$ of all transistors must be probed for each TVD logic gate. Although information about the threshold voltage of transistors cannot be obtained from IC imaging or delayering, there are several dopant profiling techniques for measuring channel doping, such as spreading resistance profiling~\cite{de1998cross} and scanning capacitance microscopy~\cite{williams1999two}~\cite{sommerhalter1999high}. To prevent the aforementioned attacks and to protect the design against reverse engineering during or after fabrication, the proposed ISFET-TVD gate has been developed.
The main contributions of this paper are as follows:
\begin{itemize}
    \item Implementation of a reconfigurable ISFET-TVD gate using different solvents for different Boolean functionality.
    \item Introduction of an obfuscation scheme that allows post-manufacture programming of the gates.
    \item Demonstration of different logic gates using ISFET-TVD, evaluated in \SI{45}{\nano\meter} CMOS technology.
\end{itemize}

The remainder of the paper is structured as follows. Section II discusses the operation and basic design of the conventional TVD logic family as proposed in~\cite{7495587}. In Section III, by taking advantage of emerging technology such as ISFET, we described the proposed ISFET-TVD logic gate. Section IV provides a comparison between the proposed design and the conventional TVD. The conclusion is presented in Section V.
\begin{table}[!b]
\vspace{-0.3cm}
\caption{ TVD logic overhead versus standard CMOS gates ~\cite{7495587}.}
\centering
\label{tab:tvdvscmos}
\begin{tabular}{c|c|c|c}
%\hline
        & Delay (\%) & Power (\%) & Area (\%) \\ \hline
TVD-XOR & 10         & 9          & 80        \\ \hline
TVD-AND & 70         & 10         & 160       \\ \hline
TVD-OR  & 37         & 10         & 160       
\end{tabular}
\end{table}

\section{Conventional TVD logic family}
\begin{figure}[!t]
%{0.5\columnwidth}
   \centering
   \includegraphics*[width=0.65\columnwidth]{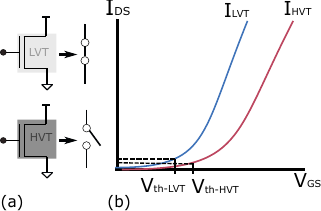}
   \caption{$V_{th}$ modulation technique: (a) $V_{th}$ programmable switch; LVT: ON, HVT: OFF, (b) IV characteristic of LVT and HVT transistors.}
   \label{fig:lhvt}
\end{figure}

The $V_{th}$ modulation technique (combining different voltage thresholds in a circuit) is widely used in the semiconductor industry to compromise between power, performance, and robustness~\cite{hamzaoglu20141}. Based on the imposed $V_{th}$ the transistors can conduct or not. For example, the transistor conducts when low $V_{th}$ (LVT) is assigned and stops conducting when high $V_{th}$ (HVT) is assigned during fabrication (Fig.~\ref{fig:lhvt}(a)). Hence, for transistors of the same size based on HVT or LVT, the amount of current conduction between the transistors can be defined as $I_{LVT}> I_{HVT}$ (Fig.~\ref{fig:lhvt}(b)). Such a characterization can be used to implement \textit{different Boolean functions} while having the \textit{same circuit structure} with different $V_{th}$ implants.

\begin{figure}[!t]
   \centering
   \includegraphics*[width=0.9\columnwidth]{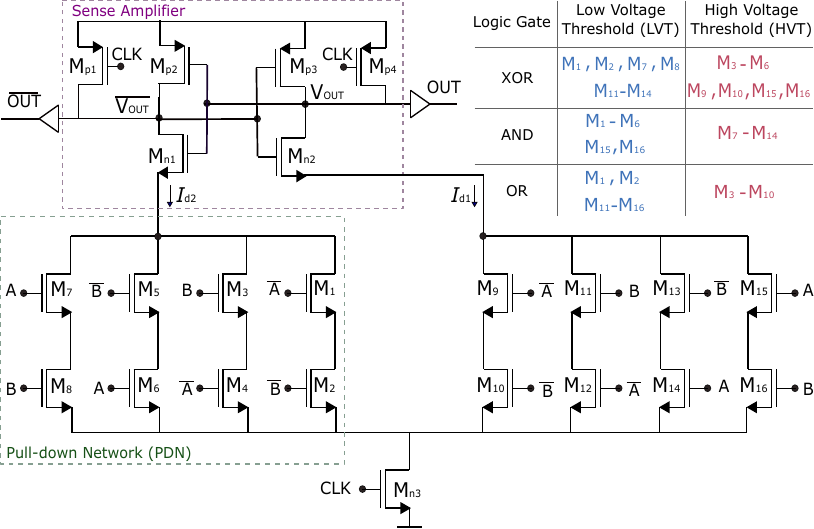}
   \caption{Conventional TVD logic family with 2-inputs (A, B); pull-down transistors considered with all possible input combinations; different Boolean functions can be realized by considering different arrangements of transistors with low or high threshold implants (LVT or HVT), described in the table.}
   \label{fig:TVD}
 %\vspace{-2.5mm}
\end{figure}

The TVD technique is one-time mask programmed with different threshold implants to realize different camouflaging gates on the same physical structure. Hence, stacks of transistors, including low (\(\sim \)\SI{300}{\m\volt}) and high (\(\sim \)\SI{600}{\m\volt}) $V_{th}$ are used as a pull-down network (PDN) to provide all possible input combinations (Fig.~\ref{fig:TVD}). A differential PDN pair is replaced by the differential input of the sense amplifier. After the gates are fired, the sense amplifier amplifies the corresponding current difference to the full logic level. The 2-input TVD logic family (A, B) is shown in Fig.~\ref{fig:TVD}. 

There are two modes of operation: when the clock is low, the circuit is in precharge mode, and when the clock goes high, the gate evaluates. While the clock is low, transistors $M_{p1}$ and $M_{p4}$ are on, so $V_{OUT}$ and $\overline{V_{OUT}}$ are high, therefore $OUT$ and $\overline{OUT}$ are pulled down. On the other hand, during the evaluation phase ($CLK$: High, $ M_{n3}$: ON), based on the input combination, only one of the parallel branches from each side of the differential PDNs is turned on. Since the transistors with the LVT or HVT are placed in an asymmetric manner, both branches conduct, but asymmetrically in the differential pairs. Therefore, the current drawn by one of the differential sides (the branch with the LVT) is greater than the other, defining the output as low or high.
The table in Fig.~\ref{fig:TVD} describes the transistors set up in differential PDNs. These must be programmed as shown using LVT or HVT implants to implement different Boolean functions.

For example, the operation for a 2-input TVD-XOR gate is as follows. The precharge phase ($CLK$: low) has already been described. After CLK is set to high (evaluation phase), for the input combination $A/B = 00$ ($\overline{A}/\overline{B} = 11$), only two branches with $M_{1}$ and $M_{2}$ (LVT) from one PDN and $M_{9}$ and $M_{10}$ (HVT) from the asymmetric side are conducting. Since $I_{LVT} > I_{HVT}$, more current will flow through $M_{1}$ and $M_{2}$, thus the $\overline{V_{OUT}}$ quickly drops ($\overline{V_{OUT}} = 0$), but $V_{OUT}$ makes a slight dip and stays high ($V_{OUT} = 1$). After the inversion, $\overline{OUT}$ goes high and $OUT$ remains low. Table~\ref{tab:tvdvscmos}~\cite{7495587} describes the overheads of the TVD logic family compared to standard CMOS gates. 

Accordingly, the same topology is used to implement other gates. This TVD logic family can camouflage the gate by using the same physical layout and design except for the threshold implant of the devices. However, the $V_{th}$ is not directly detectable by solely observing the physical layer and delayering. Still, there are several methods to measure the channel doping and ultimately reveal the gate functionality~\cite{williams1999two}~\cite{sommerhalter1999high}.
\section{Reconfigurable ISFET-TVD Logic Gate}
This section introduces the TVD gate that uses ISFET devices (ISFET-TVD) instead of MOSFET transistors with different threshold implants. The proposed camouflaging gate can be programmed after fabrication to perform different Boolean operations, while having the same physical layout and design's schematic for all logic gates.
\subsection{Emerging Technologies, ISFET}
\begin{figure}[!t]
   \centering
   \includegraphics*[width=\columnwidth]{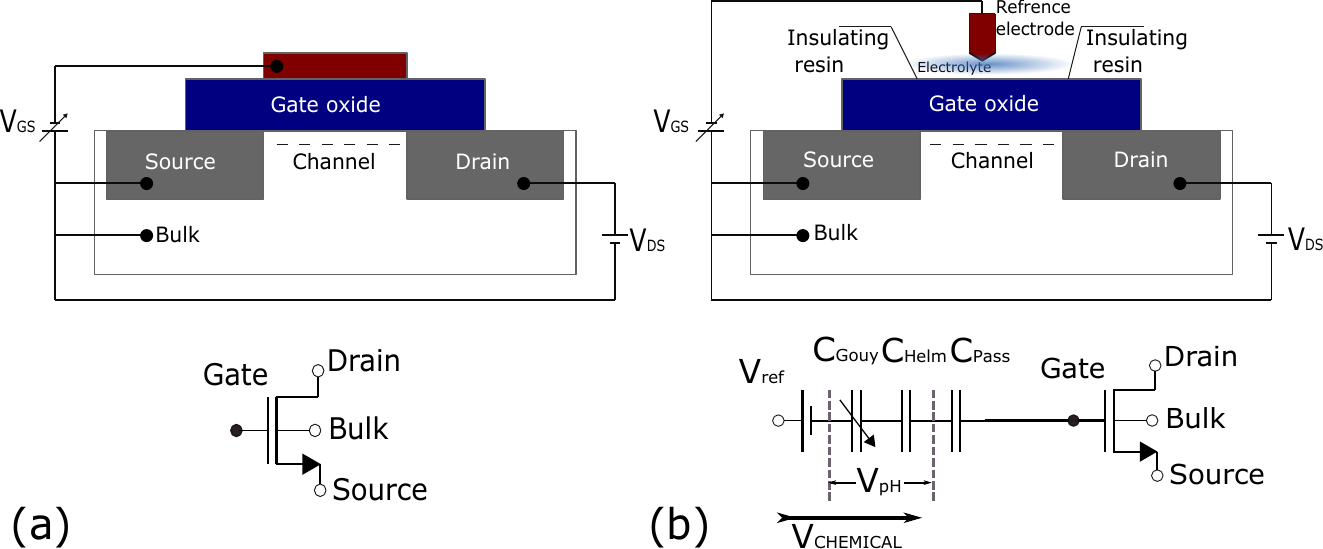}
   \caption{Schematic of a field-effect transistor (a) MOSFET and an (b) ISFET.}
   \label{fig:isfet2}
 %\vspace{-2.5mm}
\end{figure}

The ion-sensitive field-effect transistor (ISFET) operates similarly to the metal oxide field-effect transistor MOSFET. However, the gate is extended by a passivation layer that is in contact with an external reference electrode \ch{(Ag/AgCl)} through an electrolyte~\cite{moussavi2022phgen}\cite{tintelott2022realization}. Fig.~\ref{fig:isfet2} illustrates the schematics of a MOSFET (a) and an ISFET (b). For the ISFET the reference voltage is provided by an electrode, which acts as the gate voltage ($V_{G}$). Compared to conventional MOSFET structures, the threshold voltage of an ISFET ($V_{TH(ISFET)}$) device also depends linearly on the surface potential of the oxide-electrolyte interface. The expression for the drain-source current ($I_{DS}$) of an n-type ISFET transistor is~\cite{yusoff2013design}:
\begin{equation}
    I_{DS}=\mu_{n}c_{ox}\frac{W}{L}(V_{GS}-V_{TH(ISFET)})V_{DS}-\frac{1}{2}V_{DS}^{2}.
    \label{eq1}
\end{equation}

Equation~\eqref{eq1} shows that compared to the MOSFET, the threshold voltage of the MOSFET is replaced by $V_{TH(ISFET)}$, which depends on the ion concentration of the liquid. Therefore, the $V_{TH(ISFET)}$ can be changed and adjusted depending on the aqueous electrolyte that is electronically in contact with a reference electrode. This implies that $V_{TH(ISFET)}$ can be changed and reconfigured to HVT or LVT after fabrication, depending on the ion concentration of the sample solvent. 
\subsection{Proposed ISFET-TVD}
Instead of using different threshold implants for transistors in differential PDNs, a parallel configuration employing ISFET transistors is proposed. \textit{We present a camouflaged logic gate with the same schematic design and the identical layout for all logic gates, which can be programmed after fabrication depending on the Boolean functionality}. After fabrication, the devices' $V_{th}$ must be adjusted by choosing solvents with different ion concentrations depending on the Boolean function. The ion concentration is proportional to the $V_{th}$, so a low hydrogen ion strength electrolyte solution is used for LVT and a high hydrogen ion strength solution is used for HVT. Thus, the amount of charge depends on the concentration of certain ions present in the solution (hydrogen ion concentration or pH). This modulates the surface charge at the insulator-semiconductor interface of the ISFET, resulting in interface charge densities. Therefore, the pH response of an ISFET device can be characterized as a threshold voltage shift when the pH of the injected solution is varied (from 1 to 14). 

Fig.~\ref{fig:pH}(a) shows the $V_{th}$ of an ISFET device as the pH is increased from 1 to 7~\cite{narang2016ph}. In particular, for a single ISFET device, the threshold voltage increases or decreases in accordance with the pH of the device. The $I_{DS}$ characteristics are shown in Fi.~\ref{fig:pH}(b) for a fixed $V_{DS}$ and different pH values. The $I_{DS}$ is higher at lower pH values (lower $V_{th}$); e.g. $I_{DS-pH4.9} > I_{DS-pH9.2}$. Based on the Nernst sensitivity limit, the threshold voltage shift for conventional ISFET devices is \SI{59}{\m\volt}/$pH$~\cite{parizi2012exceeding}. 

Note that instead of having a transistor whose threshold voltage variation is a function of a transistor's geometry and doping, $V_{th}$ of the ISFET depends on the pH of the sample solution. The LVT or HVT can be easily adjusted after fabrication by changing the pH of the solvent in contact with the transistor. Furthermore, ISFET devices have a gate structure that is compatible with the CMOS manufacturing process. The proposed ISFET-TVD gate replaces the stack of parallel n-type transistors that have LVT or HVT threshold implants with ISFET devices in the PDN of the TVD logic family. Different Boolean functions can be implemented using the same schematic while injecting {\textit{two different pH solutions} to perform LVT or HVT (low or high pH value, respectively). With the ISFET-TVD, the gates can be reconfigured at any time after manufacturing.
\begin{figure}[!t]
   \centering
   \includegraphics*[width=\columnwidth]{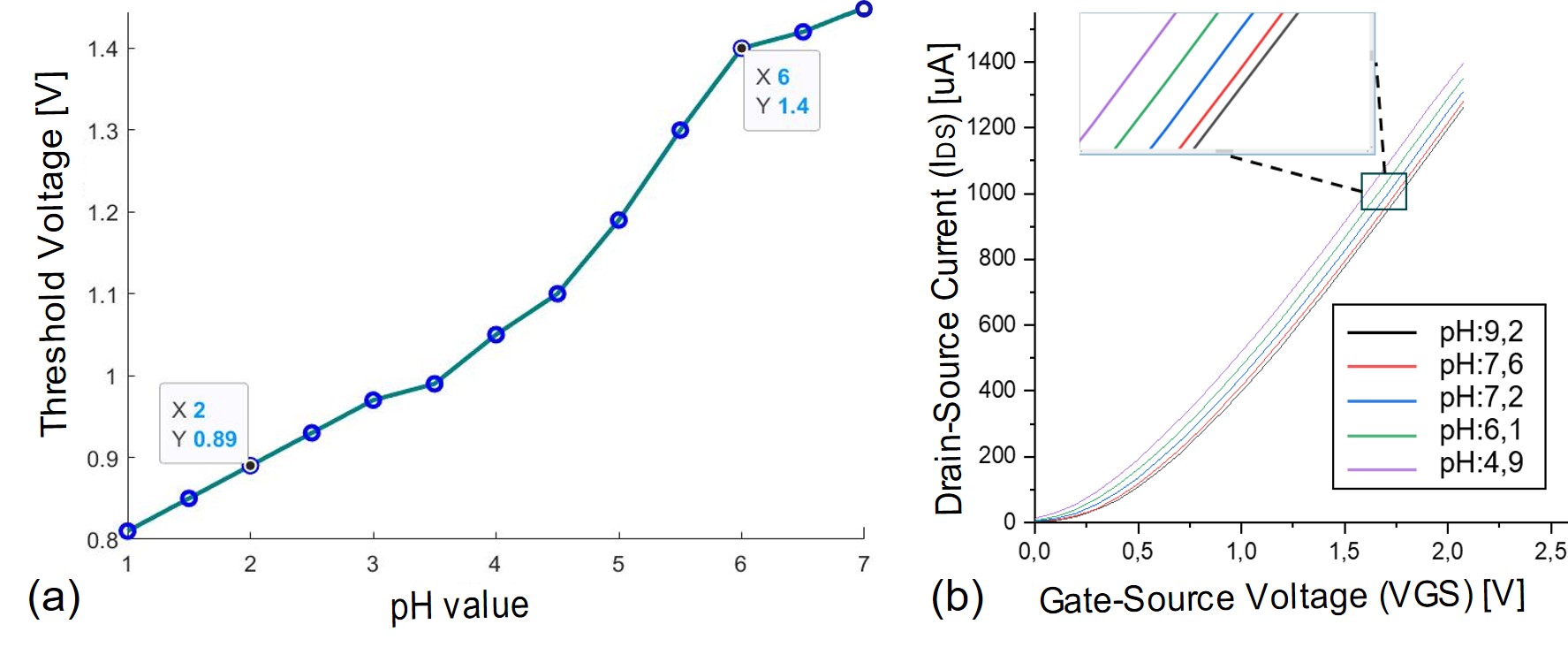}
   \caption{(a) Threshold voltage variation with respect to pH value when drain-source voltage ($V_{DS}$) is \SI{0.1}{\volt} and channel thickness ($t_{si}$) is \SI{50}{\nano\meter}~\cite{narang2016ph}; (b) drain-source current for different hydrogen ion concentrations (pH) at electrolyte interface for a fabricated ISFET device having channel width of \SI{15}{\micro\meter}, channel length of \SI{7}{\micro\meter}, and $V_{DS}$ is \SI{2}{\volt}.}
   \label{fig:pH}
% \vspace{-2.5mm}
\end{figure} 

\section{Experimental Result and Comparison}
To demonstrate the correct functionality of the new camouflaging scheme, we simulated the ISFET-TVD logic gate with 2 inputs and n-type ISFETs as it shows in Fig~\ref{fig:tvdisfet}. The conventional TVD logic family was evaluated in \SI{65}{\nano\meter} CMOS technology and the proposed ISFET-TVD was evaluated in \SI{45}{\nano\meter} CMOS technology and Verilog-A to model the surface potential, reference electrode, and electrolyte of the ISFET. All logic gates are designed using the same geometry of NMOS and PMOS devices. A supply voltage of \SI{1.8}{\volt} and a temperature of 27\textdegree{}C, and the frequency of \SI{20}{\mega\hertz} are the nominal conditions for the simulation. Cadence Virtuoso is used for simulation and analysis.

The overall design with the proposed configuration is done in the following order. First, two high and low-pH solutions are considered to inject on the ISFET devices in the differential PDNs. The pH solutions are then applied to the selected set of ISFETs according to the desired Boolean function. The input combinations are provided to the reference electrode of the ISFETs, which is electrically connected to the gate surface via the pH sample. The sense amplifier and operating phases are then used similarly to the conventional TVD design. 

The number of ISFET transistors required is the same as for conventional TVD transistors. However, a single design schematic can be used to implement different gates. The ISFET-TVD can be considered \textit{a reconfigurable universal gate} that can operate as any Boolean function based on a specific configuration of ISFETs with pH solvents. Fig~\ref{fig:result} shows the transient analysis of a 2-input ISFET-TVD configured to operate as an XOR gate for inputs A and B. 

From the results, we can observe that the ISFET-TVD behaves in the same way as the conventional TVD logic family. Therefore, in the precharge phase, $CLK = 0$ and the p-type transistors in the sense amplifier are turned on, pulling $V_{OUT}$ and $\overline{V_{OUT}}$ to VDD and thus $OUT$ and $\overline{OUT}$ to VSS. On the other hand, for the evaluation phase ($CLK = 1$), depending on the combination of inputs, only one of the branches from each side of the differential PDNs is turned on. The use of asymmetric $V_{th}$ causes more current to flow to one side of the differential PDNs, namely to the branch with LVT (low pH value), compared to its complementary branch with HVT (high pH).
\begin{figure}[!t]
   \centering
   \includegraphics*[width=\columnwidth]{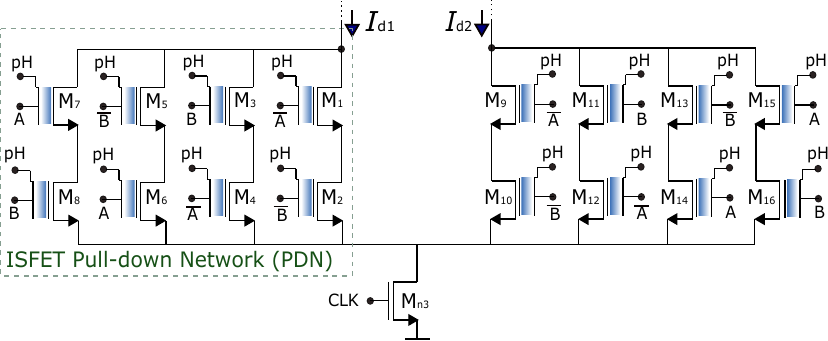}
   \caption{2-input (A,B) ISFET-TVD gate Schematic. Only differential PDNs with n-type ISFETs are shown. Low and high pH values must be placed on the ISFETs to operate as an LVT or HVT for different Boolean functionality.}
   \label{fig:tvdisfet}
\end{figure} 

Compared to the conventional TVD logic family, the proposed ISFET-TVD does not require transistors with different threshold implants; the voltage thresholds are programmed after fabrication. In addition, the ISFET-TVD gate provides the same schematic and layout for all types of logic gates. Hence, different logic gates can be programmed post-fabrication by reconfiguring the ISFET devices with different pH values.

However, the complexity of ISFETs makes it challenging to achieve an accurate, fast, and repeatable response, resulting in additional overheads. The ISFET-TVD area consumption is the same as conventional TVD~(Table.~\ref{tab:tvdvscmos}) except for the additional reference electrodes (depending on the size of the reference electrode). There is also a relatively small delay based on the time it takes for the ISFET to sense the pH (depending on the passivation layer). Furthermore, additional power is initially required to allow the solvent to flow to the ISFETs to configure/reconfigure the gate.

ISFET is an emerging technology, and there are a number of research efforts underway to overcome these challenges and to demonstrate a rapid response. On the other hand, there are some other TVD logic techniques, including enhanced-TVD (E-TVD) to reduce the amount of required transistors to enhance the overall performance~\cite{9090183}. This is important because, in the TVD logic structure, the number of transistors increases significantly with the number of inputs; this can reduce the efficiency of the gate. The proposed design has the same goals as the previous TVD designs, but uses ISFET technology to hide the design intent.

\section{Conclusion}
Reverse engineering reveals the functionality of the chip by effectively determining the gate-and-wire layout of the circuit. In this work, we proposed an ISFET-TVD camouflage gate to hide the functionality of the gate and make it RE-resilient. The ISFET-TVD extends conventional TVD logic gate technology with ISFETs. The advantage of the ISFET-TVD is that it does not require additional threshold implants and can be configured and reconfigured after fabrication. Moreover, it contains the same schematic and physical layout for all gate circuits. The gate is defined by the injected electrolyte solvent. By replacing some of the conventional gates with ISFET-TVD logic gates, the circuit can be obfuscated to further enhance the security. ISFET transistors are CMOS compatible. However, due to the complexity of ISFETs compared to conventional MOSFETs, ISFET transistors come with additional overheads. This results in a larger delay, area and power consumption compared to the conventional TVD logic family. In the future, we plan to improve the ISFET-TVD gate by reducing the number of ISFET transistors to keep the logic overhead of ISFET-TVD low enough to allow large-scale replacement of conventional gates with the proposed logic gate.
\begin{figure}[!t]
   \centering
   \includegraphics*[width=\columnwidth]{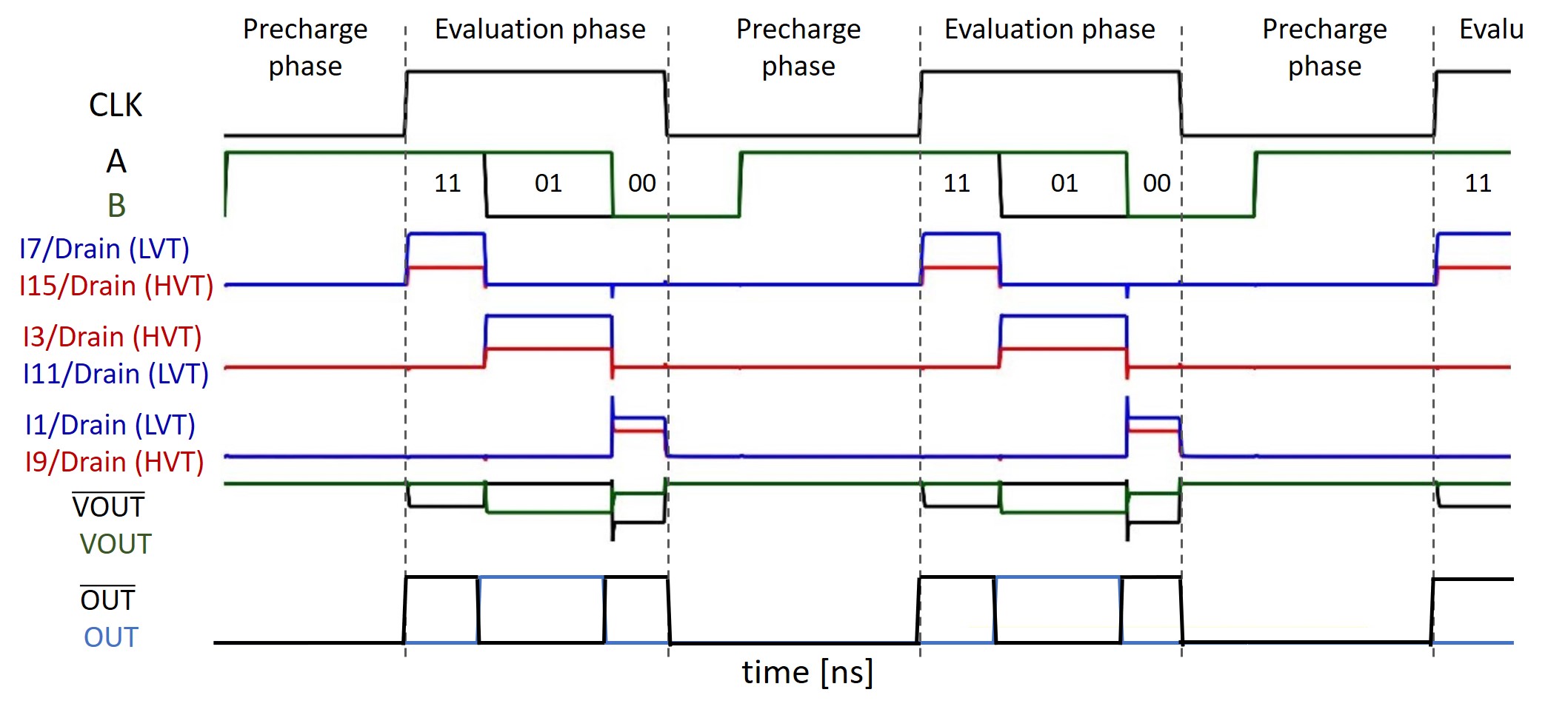}
   \caption{Transient analysis of a 2-input TVD-ISFET XOR gate for the input pH of 2 and 10, operating as devices with LVT and HVT, respectively.}
   \label{fig:result}
\end{figure} 

\section*{Acknowledgement}
This work was partially funded by Deutsche Forschungsgemeinschaft (DFG – German Research Foundation) under the priority programme SPP 2253.

\clearpage
\balance
\bibliographystyle{IEEEtran}
\bibliography{bibliography}
\end{document}